\documentclass[prl,showpacs,twocolumn]{revtex4}
\usepackage{graphics}
\usepackage{epsfig}
\usepackage{color}

\begin{document}
\title{Equilibrium strategy and population-size effects in lowest
  unique bid auctions}  \author{Simone Pigolotti$^{1,2}$, Sebastian
  Bernhardsson$^{1,3}$, Jeppe Juul$^1$, Gorm Galster$^1$, and Pierpaolo
  Vivo$^4$}

\affiliation{$^1$Niels Bohr Institute, Blegdamsvej 17, DK-2100,
Copenhagen, Denmark\\ $^2$ Dept. de Fisica i Eng. Nuclear, Universitat Politecnica de Catalunya Edif. GAIA, Rambla Sant Nebridi s/n, 08222 Terrassa, Barcelona, Spain \\
$^3$ Swedish Defence Research Agency, SE-147 25 Tumba, Sweden\\
$^4$Univ. Paris-Sud, CNRS, LPTMS, UMR8626, Orsay F-01405, France} 

\date{\today}

\begin{abstract}
  In lowest unique bid auctions, $N$ players bid for an item. The
  winner is whoever places the \emph{lowest} bid, provided that it is
  also unique. We use a grand canonical approach to derive an analytical
  expression for the equilibrium distribution of strategies. We then
  study the properties of the solution as a function of the mean
  number of players, and compare them with a large dataset of internet
  auctions.  The theory agrees with the data with striking accuracy
  for small population size $N$, while for larger $N$ a qualitatively
  different distribution is observed. We interpret this result as the
  emergence of two different regimes, one in which adaptation is feasible
  and one in which it is not. Our results question the actual possibility
  of a large population to adapt and find the optimal strategy when
  participating in a collective game.
\end{abstract}

\pacs{02.50.Le, 89.75.-k}

\maketitle

In recent years, statistical physics has provided powerful tools to
study both equilibria \cite{challet,berg} and dynamical properties
\cite{hauert, Traulsen,hanaki,frey} of games where the number of players is
large. So far, most of the efforts in this field have been focused on
games in which interactions among players are pairwise, the most
notable and studied example being the prisoner's dilemma
\cite{hauert}.  However, there are many examples of collective games
where a unique winner is singled out from a large population.  In such
cases, comparing theory with empirical data is particularly
challenging: as the number of players increases, the statistical
description becomes more appropriate. However, the complexity of the
equilibrium strategy may also increase, thus making it more difficult for
real agents to infer it from available information \cite{Baek09}.

Online auctions provide a fertile, yet scarcely investigated ground
for exploring this problem \cite{Namazi,Yang,Reichardt, Galla}: the
availability of large datasets provides a unique opportunity to study
whether strategies of real players actually maximize their winning
chances.  Here, we focus on lowest unique bid auctions. This game is
interesting for two reasons. First, it is sufficiently simple to allow
for a comprehensive mathematical analysis \cite{Baek,Ostling}.
Second, many detailed auction records are freely available online. The
game is formulated as follows: $N$ players can bid for an item of
value $V$. The bid must be a multiple of a minimum amount, so that one
can effectively consider bid amounts as natural numbers. The winner
(if any) is the player who places the lowest bid, which is unique,
\textit{i.e.} no other player bid on that amount. All players must pay a fee
$c$ to take part in the auction; additionally, the winner has to pay
the bid amount.

In this Letter, we derive an explicit analytic expression for the
symmetric Nash equilibrium of the lowest unique bid auction and
explore its dependence on the total number of players $N$. To achieve
an explicit solution we assume that $N$ is not fixed, but fluctuates
according to a Poissonian distribution, in analogy with the choice of
the grand canonical ensemble in equilibrium statistical mechanics. We
then compare the expression with a large dataset \cite{dataset} where
players are informed in advance of the total number of allowed bids
$N$. We find a remarkable change in the data as $N$ increases: in the
low $N$ regime (fewer than $\approx 200 $ players), the theory predicts
very well the bid distribution. In this regime the game is effectively
a lottery, since winning chances are evenly spread on any
number. Conversely, in the large $N$ regime, the data deviate from the
theoretical solution and rather follow an exponential
distribution. Here, players fail to adapt to the optimal strategy and
the winning chances are highly dependent on the chosen number.

In a symmetric Nash equilibrium, all players adopt the same strategy,
and no player can benefit from changing strategy unilaterally,
\textit{i.e.}, should any of the players change strategy, his expected
payoff would be equal or worse.  Consider $N$ individuals playing
according to the same strategy $\mathbf{p}=(p_1,p_2,p_3\dots)$, where
$p_k$ is the probability of bidding the number $k$. Then, the
distribution of bids is multinomial:
\begin{equation}
\mathcal{P}(\mathbf{n})=\frac{N!}{n_1!n_2!\dots}p_1^{n_1}p_2^{n_2}\dots
\end{equation}
where $\mathbf{n}=(n_1,n_2,n_3\dots)$ are the number of bids placed on each
number $k$. For convenience, we introduce the generating function:
\begin{equation}
\mathcal{G}_N(\mathbf{x})=\sum\limits_{\{\mathbf{n}\}}
\mathcal{P}(\mathbf{n})\ x_1^{n_1}x_2^{n_2}\dots
=\left(\mathbf{x}\cdot\mathbf{p}\right)^N.
\end{equation}
We now assume that $N$ is not fixed, but fluctuates according to a
Poissonian distribution of mean $\lambda$, leading to a grand
canonical generating function:
\begin{equation}
\tilde{\mathcal{G}}_\lambda(\mathbf{x})=\sum\limits_N 
\frac{\lambda^Ne^{-\lambda}}{N!}
\mathcal{G}_N(\mathbf{x})=
\exp[\lambda(\mathbf{p}\cdot\mathbf{x}-1)],
\end{equation}
\textit{i.e.} the number of bids on each number is also Poisson
distributed with mean $f_k=\lambda p_k$. By differentiating
$\tilde{\mathcal{G}}$ with respect to $\mathbf{x}$, it is
straightforward to compute the probabilities $w_k$ of $k$ being the
winning number, \textit{i.e.} that there is a unique bid on $k$ and no
unique bids on lower numbers:
\begin{equation}\label{wkeq}
w_k=f_ke^{-f_k}\prod\limits_{j=1}^{k-1}(1-f_je^{-f_j}),
\end{equation}
It is also useful to compute the probability $c_k$ of $k$ being
a potential winning number, \textit{i.e.} that there are no bids on $k$
and no winners on lower numbers:
\begin{equation}\label{ckeq}
c_k=e^{-f_k}\prod\limits_{j=1}^{k-1}(1-f_je^{-f_j}).
\end{equation}
We note that the probability of each bid to be a winner on a
given number, $w_k/f_k$, is equal to the probability, $c_k$, of numbers
to be potential winning numbers, which is a peculiar property of
Poisson games \cite{Myerson98}.

The expected payoff for a player bidding $k$ will be
$(V-k)w_k/f_k-c$. At equilibrium, the expected payoff should be
independent of $k$. Otherwise, players could benefit from bidding on
numbers with high payoffs more frequently. Imposing this condition
leads to a recurrence relation for the average bidding frequencies
\begin{equation}\label{recurs}
f_{k+1}=\ln\left(e^{f_k}-f_k\right)+\ln\left(1-\frac{1}{V-k}\right)
\end{equation}

To avoid a negative number of bids, the support of the distribution must
be limited to the region where all $f_k$'s are positive.  In
this region $k<V$ holds, so that the $f_k$'s are strictly
decreasing. The initial condition $f_1$ has to be determined
iteratively from the condition $\sum_j f_j=\lambda$.

If the frequencies $f_k$ tend to zero for values of $k$ much smaller
than the value of the item $V$ (a condition that is always fulfilled
in our dataset, and will be consistently assumed in the following),
the last term in (\ref{recurs}) can be disregarded. In this limit
(formally corresponding to $V\rightarrow\infty$) the recurrence
relation has been derived in \cite{Ostling} and it is well defined for
all values of $k\in \mathcal{N}$. We extend this solution by noting
that the normalization condition $\sum_k f_k=\lambda$ implies an
explicit expression of $f_1$: eq. \ref{recurs} can be written as
$f_k=e^{f_{k}}-e^{f_k+1}$, which summed over $k$ yields the initial
condition $f_1=\ln(\lambda+1)$. This allows for an explicit expression
for any specific $f_k$'s by iteration. Substituting the solution into
(\ref{wkeq}) also shows that the average chance of winning of each bid
is equal to $(\lambda+1)^{-1}$, which is also equal to the chance of
having no winner in the auction.

Before discussing the comparison with the data, we briefly study some
properties of the Nash equilibrium distribution. When $f_k \gg 1$, a
continuous approximation of (\ref{recurs}) shows that $f_k$ decreases
logarithmically with $k$. For small $f_k \ll 1$, one can approximate
$f_{k+1}\approx f_k^2/2$, showing that the distribution has a
super-exponential cutoff. The scaling of the value $k^*(\lambda)$,
around which the cutoff occurs, can be estimated in the following way.
By removing all the players that bid on $k=1$ we change the average
number of players by the amount $f_1$ giving
$\lambda_{new}=\lambda_{old}-\ln(\lambda_{old}+1)$.  This is
equivalent to shifting the whole distribution by one along the $k$
axis, resulting in $k^*_{new}=k^*_{old}-1$.  This scaling
transformation in the continuum limit becomes
$d\lambda/dk^*=\ln(\lambda+1)$, with the solution
$k^*(\lambda)=\mathrm{li}(1+\lambda)+C$, where
$\mathrm{li}(z)=\int_0^z dt/\ln t$ is the logarithmic integral and $C$
is a constant of $\mathcal{O}(1)$.

\begin{figure}[htb]
\begin{center}
\includegraphics[width=8.4cm]{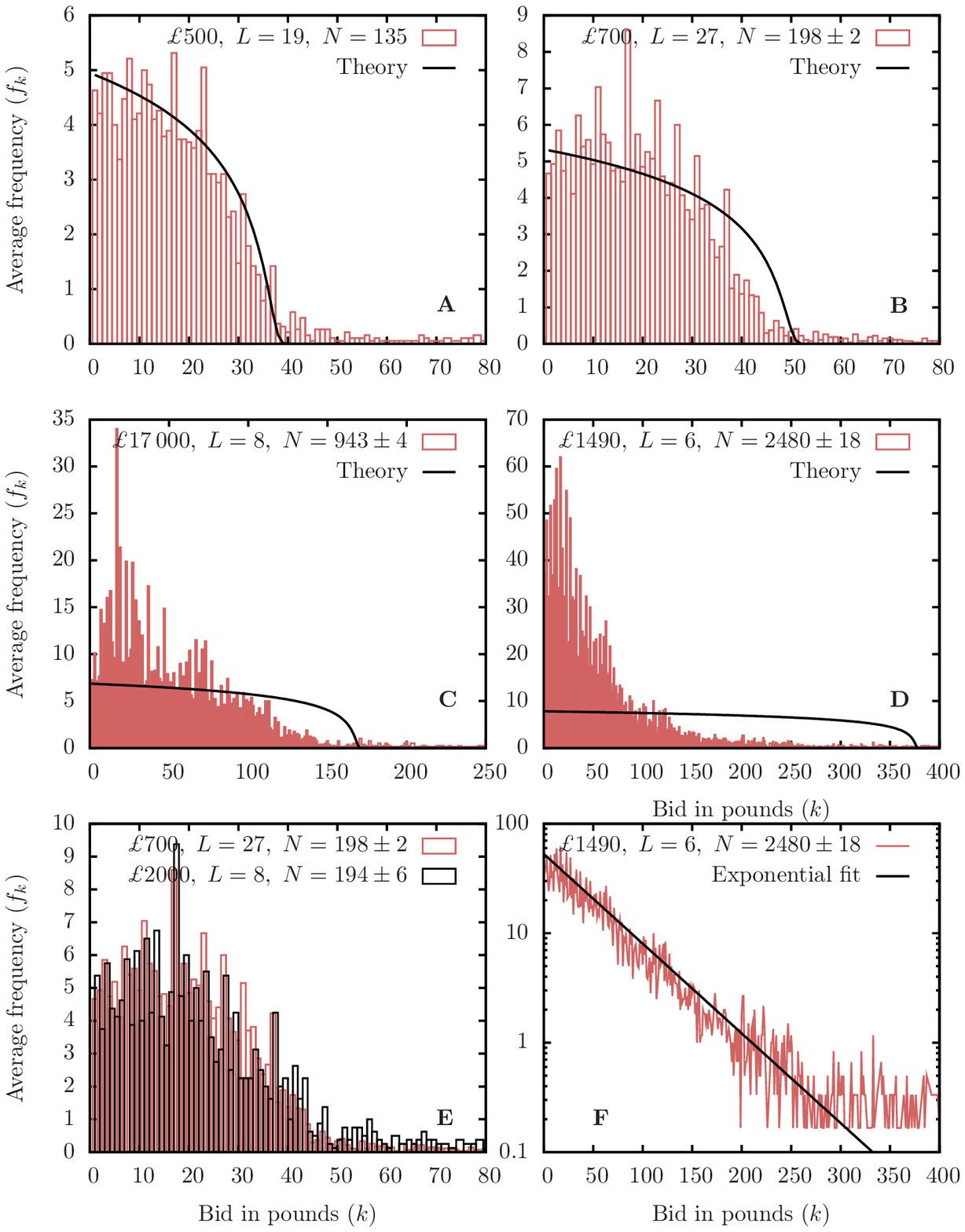}
\caption{Histograms of bidding frequencies compared with the
  theoretical equilibrium distribution. Panels A-D show average
  histograms of different auctions with similar number of players and
  same item value. Panel E compares histograms of auctions with
  similar number of players, but significantly different item
  values. Panel F is same as panel D but in linear-log scale, to
  illustrate the exponential shape of the empirical distribution at
  large $N$.\label{fig1}}
\end{center}
\end{figure}

We now move to the comparison of the equilibrium solution with
empirical data from the website \cite{dataset}. The dataset includes
$724$ online auctions from April 2007 to January 2011 with a variable
number of bids ranging from $N=26$ to $N=4748$. The number of allowed
bids in a particular auction is announced before bidding starts, and
the auction closes when this number is reached.  Each player is
allowed a limited number of bids. The average number of bids per
player in the full dataset is only $2.47$ and very weakly dependent on
$N$. In {\em Supplementary information}, we demonstrate with agent
based simulations that allowing a small number of multiple bids per
player do not alter significatively the equilibrium strategy. For this
reason, in the subsequent analysis we will neglect the effect of
multiple bids by the same player and treat the bids as statistically
independent.

In Fig.\ \ref{fig1} we compare the theoretical and empirical bidding
frequencies in different auctions having different numbers of
players and different item values. In order to make the
histograms smoother, we averaged each panel over $L$ different
auctions having similar numbers of players and same item value
(shown in the figure).

On the fine scale of single numbers, the data show a structure
dictated by known psychological effects. Players tend to favor odd
numbers over even numbers. Some specific numbers (like $17$ and other
primes) are perceived to be `original' by some players, and are chosen
with significantly larger probability than neighboring ones.

On a coarser scale, the agreement between theory and data is striking
for smaller auctions (\textit{i.e.} fewer than 200 players, panels A and B of
Fig.\ \ref{fig1}). It is particularly remarkable that the empirical
histograms reproduce the sharp cutoff, considering the non-trivial
dependence of $k^*$ on the number of players.

Theoretically, players should adjust their bidding strategies
according to the number of players rather than the item value,
which can be assumed to be infinite. In all auctions in the dataset,
the cutoffs of the corresponding theoretical distributions occur at
bid values much smaller than the item values (shown in panels). To
test whether the empirical bidding distributions are independent of
$V$, panel E compares two sets of auctions with the same number of
players, but with item values that differ by a factor of three. The
bidding distribution for the pricey item have a slightly heavier tail,
but overall the distributions are very similar.

The agreement between theory and data progressively deteriorates as
the number of players increases (see panel C and D). For a large number of
players (more than $2000$) an exponential distribution fits the
data better, as shown in the lower panels of Fig.\ \ref{fig1}. This can be
understood by arguing (see e.g.\ \cite{mckelvey,chen,goeree}) that
players having incomplete information about the game play
according to a strategy defined as $p_k=\exp(-\beta
E_k)/\sum\limits_j \exp(-\beta E_j)$
where $E_k$ is the expected payoff when placing a bid $k$ and $\beta$
quantifies the degree of uncertainty about the game. If players have
poor knowledge about the optimal strategy, they could assume a uniform
prior probability to win on each $k$, resulting in $E_k$ decreasing
linearly with $k$ due to the cost of the bid when
winning. Substituting this prior into the logit strategy yields a
strategy exponentially decreasing with bid size.

\begin{figure}[htb]
\begin{center}
\includegraphics[width=8.4cm]{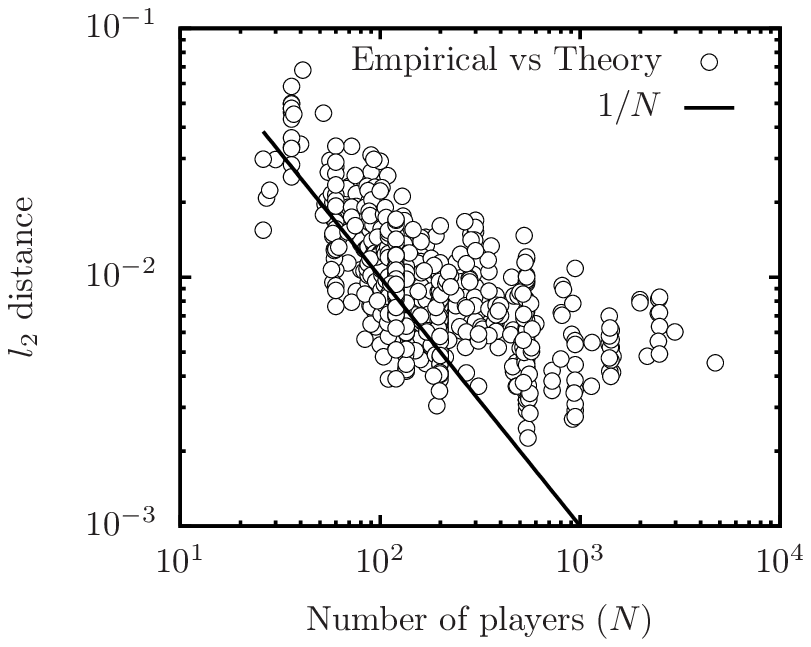}
\caption{$l_2$ distance between the theoretical solution and the
  empirical histogram for each auction. The continuous line is the
  theoretical expectation $\langle d\rangle =N^{-1}$ if all bids for
  each histogram were randomly drawn for the theoretical
  distribution.\label{fig2}}
\end{center}
\end{figure} 
A quantitative measure of how the data deviate from theory is shown in
Fig.\ \ref{fig2}, where, for each auction, we plot the $l_2$ distance
$d$ between the theoretical probability distribution and the empirical
one, $d=N^{-2}\sum_k(f_k-\phi_k)^2$, where $\phi_k$ is the number of
bids on $k$ in a given auction. If bids were randomly drawn from the
theoretical distribution, the expected distance would decrease with
the number of players as $\langle d \rangle=N^{-1}$. Empirically, the
distances have a large spread around the expected value for small
auctions and are consistently larger than expected for $N>200$. This
outcome cannot be simply explained by the larger number of auctions
for smaller $N$: as a consequence of players' turnover, the
distributions at fixed $N$ do not evolve with time in a significant
way, as we checked by comparing older to more recent auctions.

\begin{figure}[htb]
\begin{center}
\includegraphics[width=8.4cm]{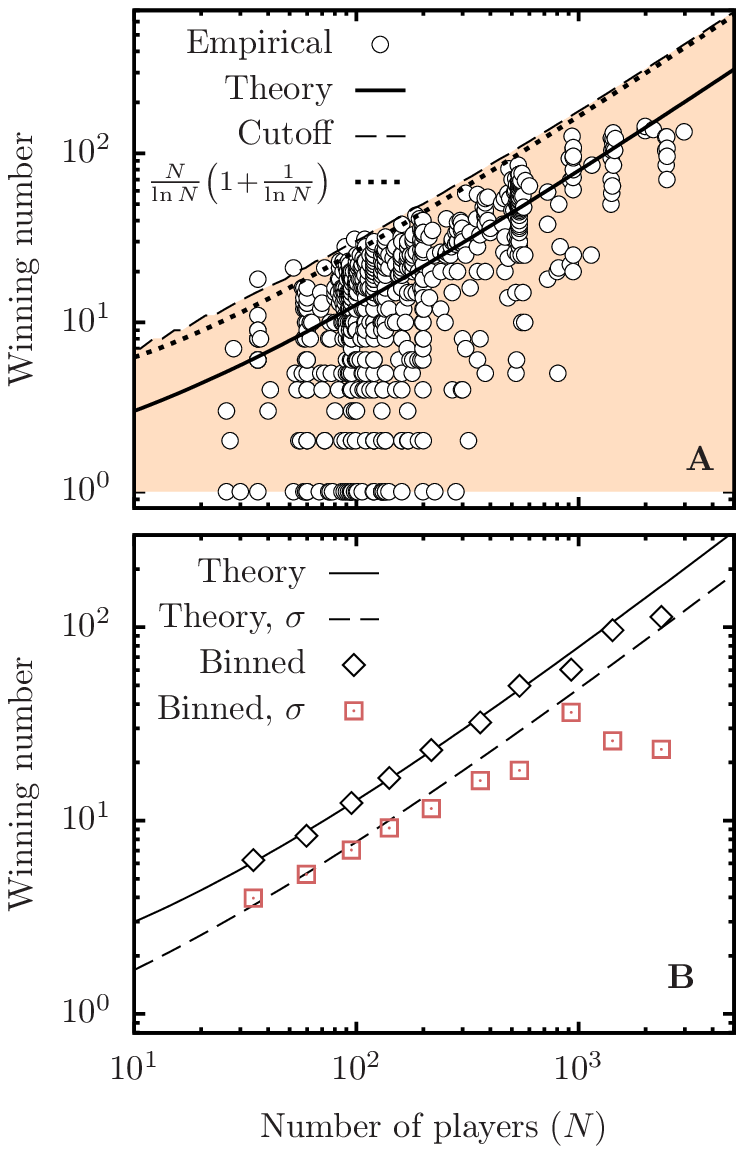}
\caption{Panel A: Winning numbers as function of auction size. The continuous
  line is the theoretical average and the shaded area denotes the
  region where bids are below the theoretical cutoff $k^*$. Panel B:
  average and standard deviation of the winning numbers, compared with
  the respective theoretical lines. \label{fig3}}
\end{center}
\end{figure}

Another interesting quantity is the distribution of actual winning
numbers. At equilibrium, it should be
equal to the distribution of bids. As shown in Fig.\ \ref{fig3}, the
empirical distribution of winning numbers supports this feature. The
vast majority of winning numbers fall within the theoretical cutoff
(pink shaded area of Fig.\ \ref{fig3}). In the figure, we also show
the analytical estimate $k^*\approx (1+1/\ln N)N/\ln N$ based on the
asymptotic expansion of the logarithmic integral. The black line is
the average winning number, which in the relevant range, scales
approximately in the same way as the cutoff.

\begin{figure}[htb]
\begin{center}
\includegraphics[width=8.4cm]{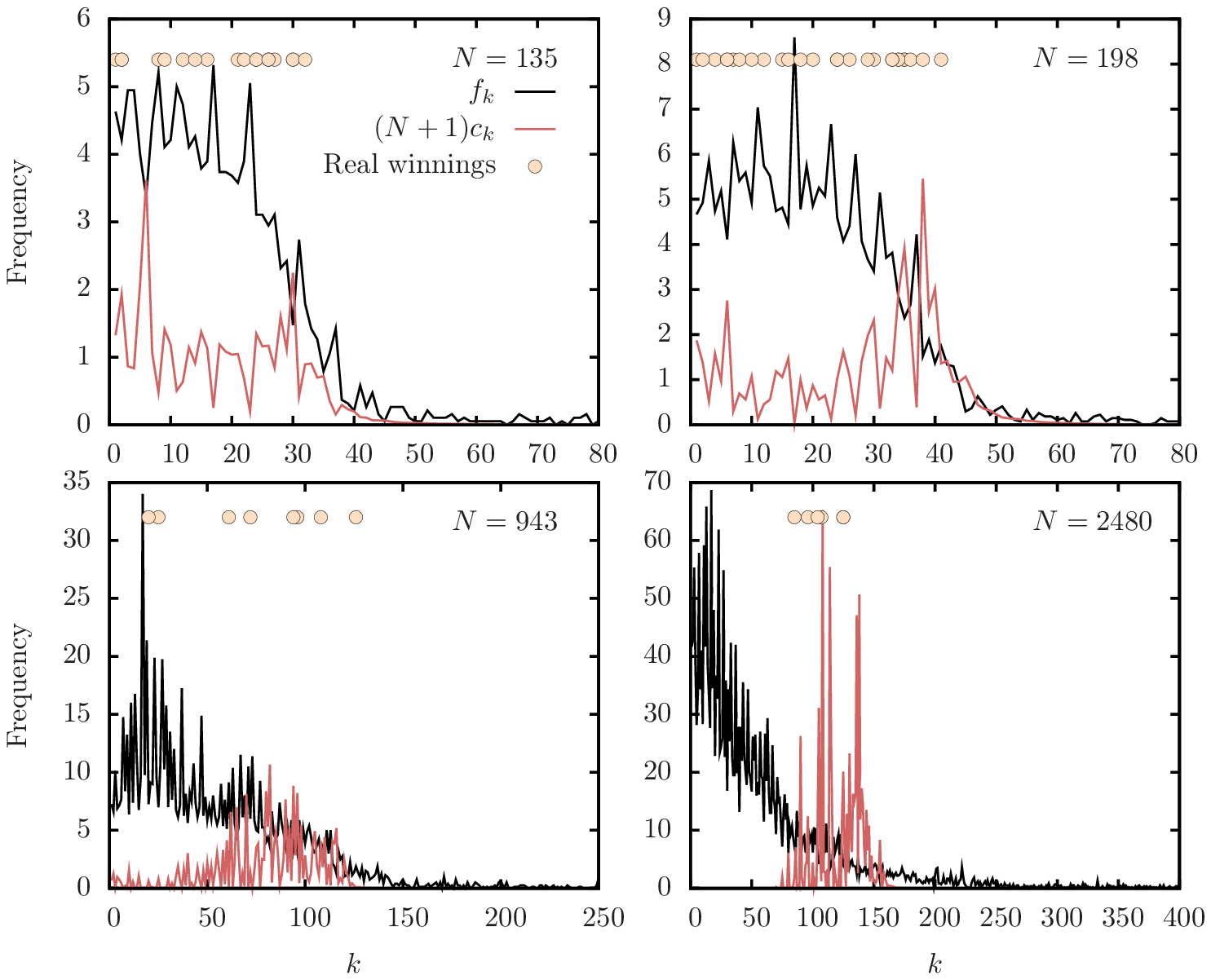}
\caption{Winning chances when bidding $k$. Black lines are the same
  empirical data as panels A-D of Fig. \ref{fig1}. Gray (red online) lines
  are the relative probabilities to win when betting on $k$ given that
  all other players bid according to the empirical distribution. 
To help the comparison, we have renormalized the winning probabilities, 
 $c_k$, with their value for the Nash equilibrium $(N+1)^{-1}$. 
  The shaded points mark the actual winning bids for each auction 
  (arbitrary y-position). 
  \label{fig4}}
\end{center}
\end{figure}

Binning the data yields average winning numbers in excellent agreement
with the theory, even for large $N$ where the empirical distribution
of bids departs from the theory. However, the variance of winning
numbers becomes much smaller than the theoretical prediction for
$N>10^3$.  To further explore this phenomenon, we compute the actual
probabilities to win on a certain number, given the empirical
distributions of bids for auctions of various sizes.  This probability
is given by $w_k/f_k$ which, since $c_k=w_k/f_k$, is the same as the
probability to win on a certain number for an additional player
entering the game. The results are shown in Fig.\ \ref{fig4} for the
same examples considered in Fig.\ \ref{fig1}. For auctions with few
players, the largest chance of winning is not more than $4-5$ times
higher than $(N+1)^{-1}$, the winning probability at the Nash
equilibrium. In this region, the game is not very different from a
lottery, as the winning chances do not depend much on the chosen
number $k\ll V$.  In contrast, for large auctions, the winning chances
on low and high numbers are very small, while intermediate bids may
have a winning chance more than $60$ times higher than the average bid
in the Nash equilibrium, providing opportunities for exploitation and
thus potential room for adaptation.

Summarizing, in this Letter we derived the analytical equilibrium
bidding distribution for the lowest unique bid auction and compared it
with empirical data. The emerging picture is that players are able to
infer the optimal strategy with striking accuracy when the number of
players $N$ is not too large. In the large $N$ regime, the
population-level strategy is highly non-optimal and it seems to be
determined by the simple principle of not assuming a particular
preference for any number while avoiding the cost of large bids.

This result raises non-trivial questions about the effectiveness of
adaptive dynamics in collective games. While it has been studied
whether adaptation eventually drives the system to the optimal
solution or generates more complex dynamical behaviors \cite{novak},
it is also crucial to assess \emph{how fast} the equilibrium is
reached. A thorough study of adaptive dynamics in such system will be
the subject of a future study; in {\em Supplementary Information}, we
show that indeed, in the simple case of replicator dynamics, adaptation
becomes slower at larger $N$. Such slowdown could prevent the
inference of the equilibrium strategy for large populations in
realistic time-scales, and thus explain the increasing lack of
adaptation in our data as $N$ grows.

We acknowledge useful discussions with M. Marsili, K. Sneppen,
C. Strandkvist and P. Kempson.\\
\newpage

{\large{\bf \noindent Supplementary material for: Equilibrium strategy
    and population-size effects in lowest unique bid auctions }}

\vspace{0.5cm}

\section{Effect of multiple bids per player}

\begin{figure}[htb]
\includegraphics[width=7.5cm]{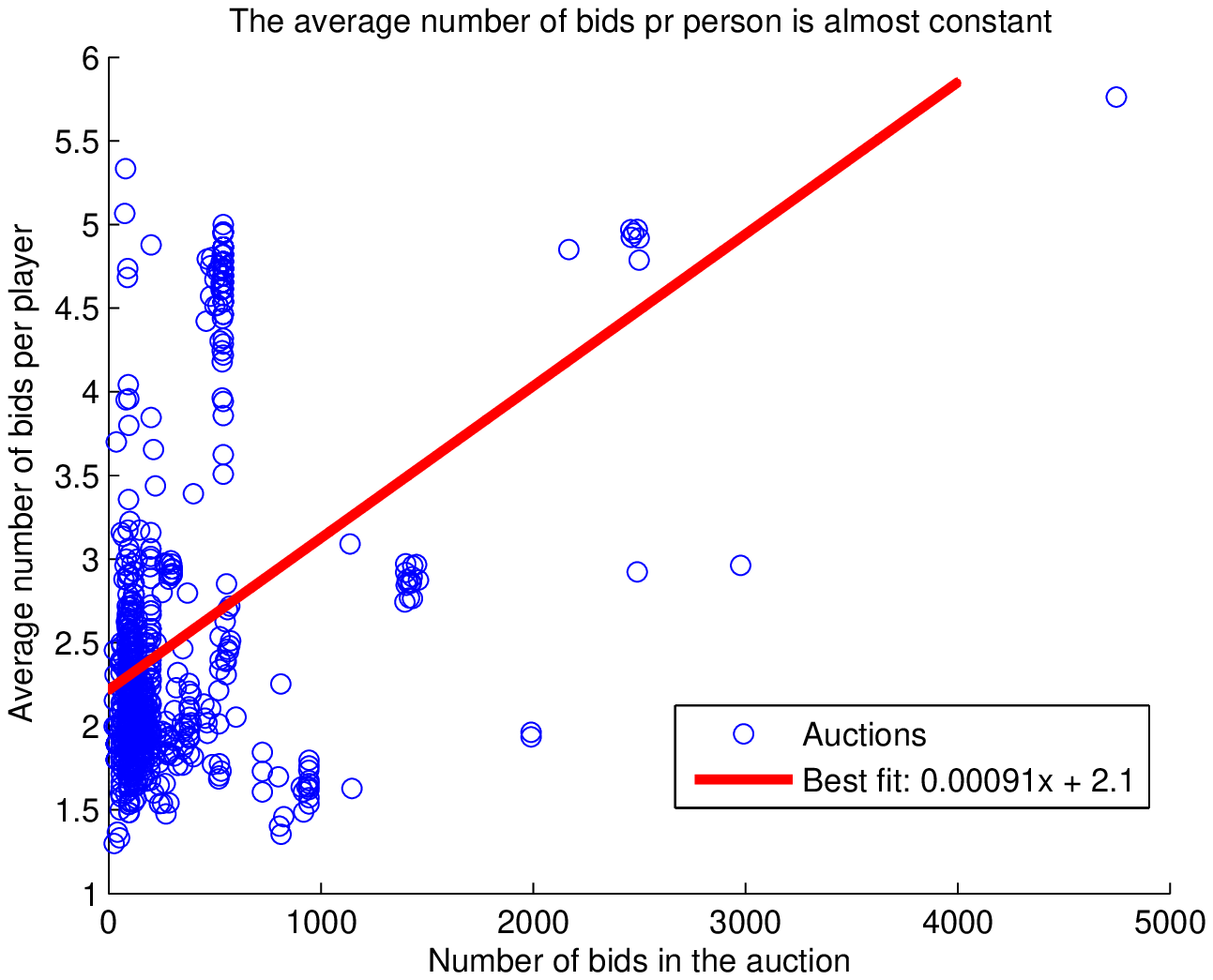}
\includegraphics[width=7.5cm]{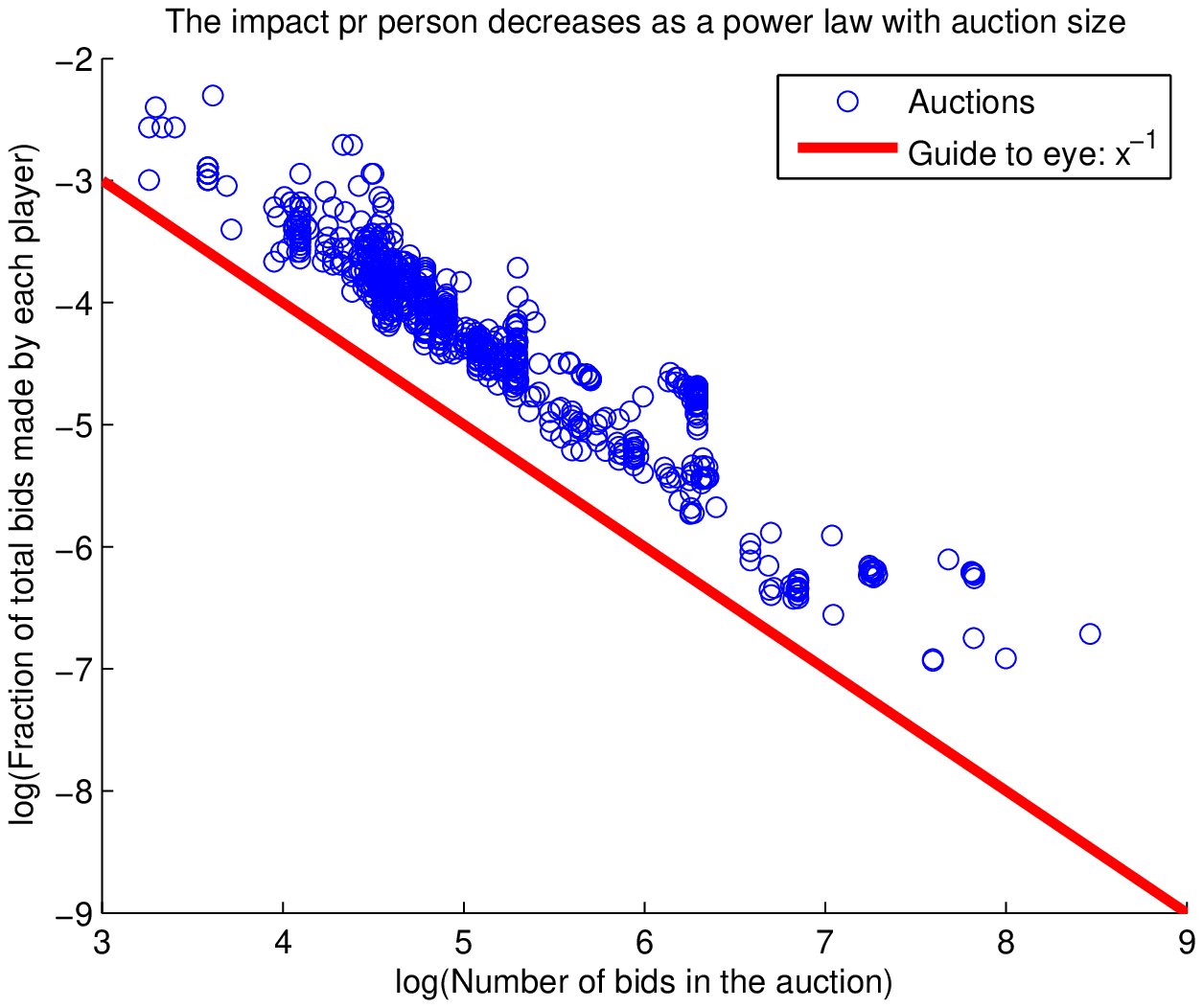}
\caption{(top) Average number of bids per player in each auction is
  only weakly depended on the number of bids in the auction.  (bottom)
  Average fraction of bids per player decreases with the total number
  of bids. \label{figS1}}
\end{figure}

To test the effect of multiple bids, we analyzed the statistics of the
number of bids per player in the dataset. As stated in the Letter, the
average number of bids per player is $2.47$. Fig. (\ref{figS1}, top)
shows how this average is weakly dependent on $N$. This means that the
impact of each player (that is, the average fraction of the total
number of bids per player) decreseas with $N$, as shown in
Fig. (\ref{figS1}, bottom). This means that the discrepancy between
theory and data as $N$ increases is very unlikely to be caused by the
effect of multiple bids per player.

\begin{figure}[htb]
\begin{center}
\includegraphics[width=6cm,angle=-90]{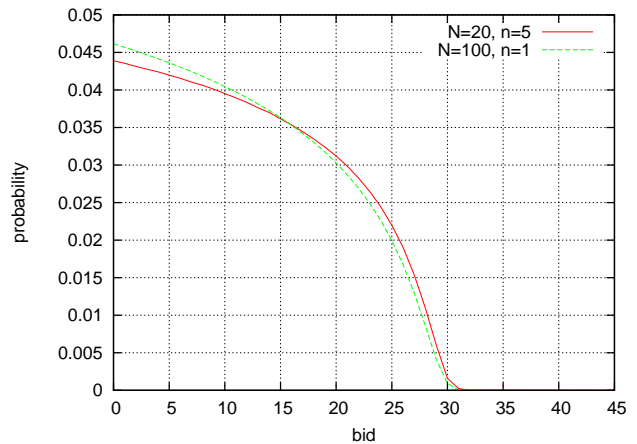}
\caption{Equilibrium bidding distribution for $N=100$, compared with the 
equilibrium distribution for $20$ players, each one bidding $5$ times.
  \label{figS2}}
\end{center}
\end{figure}

Finally, in Fig. (\ref{figS2}) we compare the Nash equlibria for auctions
of 100 bids in the two cases where $20$ players are bidding $5$ times
each (solid red line) and $100$ players are bidding $1$ times each
(dashed green line). The Nash equilibria in this version of the game
are found using individual-based simulations, where all players bid
according to the same distribution, except that they avoid placing two
bids on the same number in the same round. The distribution is then
adapted according to the statistics of winning numbers, until reaching
an equilibrium in which bids on each number have the same winning
probability. The figure shows that, even with $5$ bids per player, the
equilibrium in this version of the game is very resemblant to the
solution calculated in the Letter, where all bids are independent.

\section{Replicator Dynamics}
In this section, we study the dependence of adaptation speed on the
number of players $N$. As the details of the outcome might depend on
the specific adaptation rules, we present here results obtained by
replicator dynamics, which is one of the most general forms of
macroscopic equations for adaptive dynamics. Indeed, one can show that
many microscopic implementations of adaptive dynamics either converge
macroscopically to the replicator equation or to its modified form
\cite{Traulsen}.

\begin{figure}[htb]
\begin{center}
\includegraphics[width=8cm]{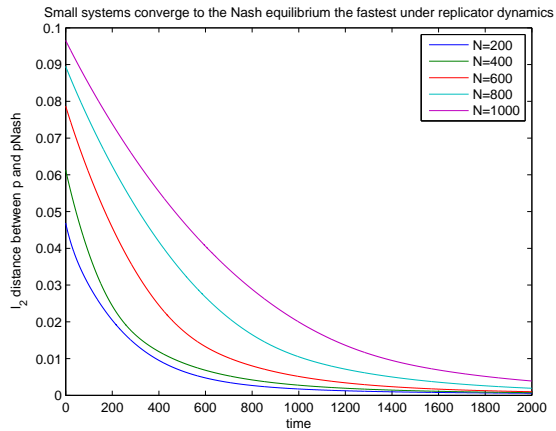}
\caption{Distance from the Nash equilibrium of replicator dynamics,
  Eq. (\ref{repl}), as a function of time (in arbitrary
  units). Different curves are for different values of
  $N$.\label{figS3}}
\end{center}
\end{figure}

Replicator dynamics is defined by the system of equations:

\begin{equation}\label{repl}
  \frac{d}{dt}p_k=p_k\left[c_k(\mathbf{p})-\sum_{i=1}^\infty 
p_i c_i(\mathbf{p})\right]
\end{equation}

where the $c_k$ are the expected payoffs of each strategy $k$, which
depend on the state of the system and are determined by Eq. (5) in the
Letter. We chose as initial distribution $p_k(t=0) = A \exp(-k/30)$,
$A$ being a normalization constant. Fig. (\ref{figS3}) shows that the
probability distribution converges to the Nash equilibrium slower and
slower as $N$ is increased. Similar results are obtained when
starting with different initial distributions (apart from a difference in
the initial overlap with the Nash equilibrium).

\end{document}